# Comparison of Different Spike Sorting Subtechniques Based on Rat Brain Basolateral Amygdala Neuronal Activity

Sahar Hojjatinia
School of Electrical Engineering and Computer Science
The Pennsylvania State University
University Park, USA

Constantino M. Lagoa
School of Electrical Engineering and Computer Science
The Pennsylvania State University
University Park, USA

*Abstract*— Developing electrophysiological recordings of brain neuronal activity and their analysis provide a basis for exploring the structure of brain function and nervous system investigation. The recorded signals are typically a combination of spikes and noise. High amounts of background noise and possibility of electric signaling recording from several neurons adjacent to the recording site have led scientists to develop neuronal signal processing tools such as spike sorting to facilitate brain data analysis. Spike sorting plays a pivotal role in understanding the electrophysiological activity of neuronal networks. This process prepares recorded data for interpretations of neurons interactions and understanding the overall structure of brain functions. Spike sorting consists of three steps: spike detection, feature extraction, and spike clustering. There are several methods to implement each of spike sorting steps. This paper provides a systematic comparison of various spike sorting sub-techniques applied to real extracellularly recorded data from a rat brain basolateral amygdala. An efficient sorted data resulted from careful choice of spike sorting sub-methods leads to better interpretation of the brain structures connectivity under different conditions, which is a very sensitive concept in diagnosis and treatment of neurological disorders. Here, spike detection is performed by appropriate choice of threshold level via three different approaches. Feature extraction is done through PCA and Kernel PCA methods, which Kernel PCA outperforms. We have applied four different algorithms for spike clustering including K-means, Fuzzy C-means, Bayesian and Fuzzy maximum likelihood estimation. As one requirement of most clustering algorithms, optimal number of clusters is achieved through validity indices for each method. Finally, the sorting results are evaluated using inter-spike interval histograms.

## I. Introduction

One of the principal components of the brain's microscopic structure is the neuron cell. The importance of this concept has led to large research efforts to understand the mechanism behind neuronal activity. Neurons interact with each other by receiving and sending electric action potentials or spikes. From the neuronal level, we can go up to the neuronal circuits, cortical structures, the whole brain, and finally to the behavior of the organism [1]. So, it is important to recognize the activity of each individual neuron from electrophysiological recordings.

Analysis of electrophysiological data recordings from single neurons activity plays an important role in understanding the brain function. Recorded signals are always contaminated by high amounts of background noise from either the recording system or the activity of farther neurons. Moreover, collected data is a combination of activity of several neurons near the recording site that might be too intricate to interpret without any preprocessing [2].

Understanding the brain function requires neuronal activity decoding within networks of neurons [3]. Spike sorting is an interpretation tool to analyze the massive neural recordings and uses to identify individual cell activity in electrophysiological recordings. This process facilitates discerning singular neuronal activity while some neurons fire concurrently [2]. The derived information from spike sorting improves the ability of neuroscientists to interpret the meaning of experiments as well as elevating effectiveness of potential neurological treatments and interventions.

Spike sorting is the process of isolating spikes from the background noise, extracting prominent features from the detected spike waveforms, and correctly assigning each spike to its originating neuron [4]. This process can be done by an appropriate choice of spike sorting sub-techniques. There are several automatic, manual, and semi-automatic methods to conduct spike sorting, each with its own complexity, advantages, and disadvantages [5].

Applying conventional spike detection technique, spikes are usually separated from noise by manual amplitude threshold discrimination. In this method, the threshold is assigned based on the elements related to distribution of background activity amplitude and true spikes amplitude [2]. Threshold level can also be applied automatically, based on the standard deviation and median of the signals.

Spikes from different neurons can be distinguished by their distinct features from the others. After applying the threshold and detecting the spikes, the next step is to extract critical feature coefficients from the detected spikes. There are several methods to extract spike features, with principle component analysis (PCA) being one of the most commonly used ones [6]. PCA converts a high dimensional data to a lower dimension feature space containing linearly correlated data. One limitation of PCA is that it does not always produce several distinguishable clusters [8].

PCA has been extended by scientists for various industrial purposes [7]. These linear extensions consist of multiway PCA, dynamic PCA, recursive PCA, etc. Despite their improved performance than the standard PCA for specific tasks, these linear projection methods fail to explain the nonlinear interactions among signals. To address this limitation, some nonlinear PCA-based methods have been proposed [8]. Still, the new techniques had difficulties in handling nonlinear optimization problems. To solve the issue, kernel PCA (KPCA) technique has been proposed and applied for nonlinear process monitoring [9]. KPCA provides the opportunity of computing principle components in high dimensional feature spaces using the kernel trick [10].

The final step in spike sorting is to cluster the featured spikes via applying a clustering algorithm and assign each spike to its originating cluster. There are different clustering algorithms to be applied. Some of these techniques consist of artificial neural networks [11]–[13], K-means clustering [14], Fuzzy C-means [15], hierarchical clustering [16], Gaussian mixture model [2], t-distribution mixture model [17], Bayesian clustering [18], and Fuzzy maximum likelihood estimation [19], etc.

To evaluate the clusters, clustering validation methods, known as validity indices, are required. Validity indices are extremely important to determine the optimal number of clusters. One approach to discern the best number is to iterate the whole clustering process by applying a minimum and maximum number of clusters, and assess the results via validity indices. There are different methods to evaluate each of clustering algorithms. The Xie-Beni index, XB, calculates the overall average compactness versus separation of the clusters [20]. Another index is Calinski and Harabasz index, or in short CH index [21]. The fuzzy hyper volume index, $V_{FH}$, which is introduced by Gath and Geva employs the concepts of hyper volume and density [19].

To best of our knowledge, no study has performed a systematic comparison between analytical results of applying available spike sorting sub-techniques for spike detection, feature extraction and clustering, considering real neuronal activity data. Exploring the effectiveness of different tools and their drawbacks based on brain real electric signals paves the way for a more reliable choice of these techniques by neuroscientists. To address this shortcoming, the main objective of this paper is to identify the appropriate spike sorting sub-methods which will result in more accurate data analyses and consequently better representation of brain connectivity and nervous system function. In this regard, for spike detection, we apply threshold via three methods consist of empirical technique, median-based approach, and standard deviation-based method. Then, PCA and KPCA are employed for feature extraction. Afterwards, various clustering algorithms such as K-means, Fuzzy C-means, Bayesian and Fuzzy maximum likelihood estimation are applied to the featured data and the succeeding outcomes are shown. Three methods consist of XB, CH, and $V_{FH}$ are used for determining the optimal number of clusters.

## II. METHODS

### A. Overall process

Spike sorting is a clustering technique aimed at identifying the activity of single neurons in electrophysiological recordings. In this paper, we employ real data from the basolateral amygdala (BLA) of rat brain to apply the sorting analysis. The procedure for spike sorting can be abbreviated in three steps: spike detection by filtering and applying threshold, extracting waveform features, and spike clustering which each will be explained in the following.

### B. Data source

Extracellular firing activity data was recorded from the BLA of male Wistar rat brain [22]. The rat was housed in Animal Care Facility maintained at 23 ± 1∘C on a 12:12 h light/dark cycle. Food and water supplied ad libitum. All procedures performed according to the Guide for the care and use of laboratory animals (National Institutes of Health Publication No. 80–23, revised 1996) and were approved by Research and Ethics Committee of Shahid Beheshti University of Medical Sciences, Tehran, Iran.

Extracellular recording from individual neurons obtained with tungsten microelectrode (shaft diameter 127 μm, tip exposure 1–3 μm, tip impedance 5 MΩ; Harvard Apparatus, Holliston, MA). Using stereotaxic instrument, microelectrode advanced into the BLA (-2.52 mm posterior to bregma, -4.8 mm lateral to midline, and -8.4 mm ventral to skull surface) according to rat brain atlas [23]. Signals were recorded using a data acquisition system, filtered between 300 and 10000 Hz, and sampled with the rate of 50 kHz (D3109; WSI, Tehran, Iran). Each recording lasted for 30 minutes. For the current research, the trial period is considered 200 s that is 10,000,000 data points to reduce the computational complexity.

### C. Spike detection

Identification of spikes in extracellularly recorded data and classification of neural activity require a first key step in the spike sorting process: action potentials detection. One of the distinct characteristics of the signals which distinguishes them from the background activity is peak amplitude. These peaks can be separated by applying an appropriate threshold. It should also be considered that in extracellular recordings, spikes are recorded as negative peaks. The reason is that the recording medium potential is negative with respect to the cell potential. Therefore, a negative threshold should be applied.

There are several methods that have been applied by researchers to discern the appropriate value of the threshold. Here, we apply three of the most used methods. First method which is mostly common among biologists is a practical technique. In this manual method, threshold assigns based on the empirical logics and background noise level. This process can be performed by Plexon offline sorter software (Plexon Inc., Dallas, TX). However, this method is affected by the experience level and judgement of the experimenter [24]. Second and third methods use the median and standard deviation of the data. In the median-based method, the threshold is set as [25]

$$threshold = k.median\left\{\frac{|x|}{0.6745}\right\} \quad (1)$$

where $x$ is the original signal including the background noise and $k$ is a constant that its value can be three to five. Threshold amplitude based on the standard deviation of neural signals can be achieved by

$$threshold = k.std(x) \quad (2)$$

where again $x$ is the neural activity consisting spikes and noise and $k$ is a constant usually between three and five. For both methods, we apply all three values of constant $k$ to achieve the optimum amount considering signal to noise ratio.

It is noticeable that no matter which method is applied, the threshold level always discerns a trade-off between the missed spikes and the adding noise which may pass that level. So, it might be somewhat complicated to decide about the optimal technique to determine the threshold level.

*D. Spike feature extraction*

Feature extraction methods determine a subspace with a dimension less than equal to the dimension of the original feature space, either in a linear or a nonlinear fashion [14]. Linear dimension reduction techniques, such as PCA which is one the well-known tools for extracting features of spikes, are widely used by scientists. PCA is a statistical procedure that employs orthogonal transformation and converts a set of observations of possibly correlated variables into a set of values of linearly uncorrelated ones while preserving as much information as possible. This process reduces the dimension of data via retaining the portion of data which has the most significant variability for data analysis. PCA uses the mean square error concept for its approximation.

Kernel PCA is one of the most used nonlinear feature extraction techniques and is a generalized form of PCA. The principal idea behind KPCA method is to map the original input space into a new high-dimensional feature space by applying a nonlinear function $\phi$ and then deal with a linear PCA to obtain the principal components [14].

In this study, we have applied both PCA and KPCA for extracting the features of the real data waveforms. Gaussian radial basis function Kernel PCA is used to evaluate PCA performance as a linear measure. It is implemented by

$$K(x,y) = exp(-\beta \|x - y\|_2^2); \beta = \frac{1}{2\sigma^2} \quad (3)$$

where $\sigma$ is a free parameter to be optimized.

*E. Clustering algorithms*

The final step of spike sorting process is to arrange the featured spikes into different clusters. Through this step, each spike assigns to its originating cluster.

*K-means algorithm*

The K-means clustering algorithm is a simple and popular technique that classifies a set of data into k number of clusters, based on the extracted features of the dataset. In this method, objects within a cluster are closer to each other than elements of other clusters in terms of normalized distance [26]. After specifying the number of clusters by the user, the algorithm applies an iterative process to group the featured spikes according to the minimum Euclidean distance. This approach defines a set of boundaries that separate the clusters [2].

*Fuzzy C-means algorithm*

One of the most popular fuzzy clustering algorithms is fuzzy c-means (FCM) [27]. The algorithm is based on the minimization of the objective function $J(X;U,V)$ through an iterative process. More precisely, it aims at solving

$$\min_V J(X;U,V) = \min_V \sum_{k=1}^{n}\sum_{i=1}^{c}(u_{ki})^m\|x_k - v_i\|^2$$

$$subject\ to\ \sum_{i=1}^{c} u_{ki} = 1\ for\ k = 1, \dots, n \quad (4)$$

$$m > 1, u_{ki} \geq 0\ for\ all\ i = 1, \dots, c$$

$$u_{ki} = \frac{1}{\sum_{j=1}^{c} \frac{\|x_k - v_i\|^{\frac{2}{m-1}}}{\|x_k - v_j\|}}, v_i = \frac{\sum_{k=1}^{n}(u_{ki})^m x_k}{\sum_{k=1}^{n}(u_{ki})^m}$$

where $n$ and $c$ are the total number of data vectors in the dataset and the number of clusters, respectively. $X = \{x_1, x_2, \dots, x_k\} \subset R^s$ is the featured data and $V = \{v_1, v_2, \dots, v_c\} \subset R^s$ is the cluster centers. $U = (u_{ki})$ is a fuzzy partition matrix expressing the belonging degree of each element $x_k$ to cluster $v_i$. The parameter $m$ is called a fuzzifier and is a hyperparameter that represents the fuzziness of clusters. It has been established that $m = 1$ represents non-fuzziness of the cluster.

*Bayesian clustering*

The process of clustering based on Bayesian algorithm is to consider a multivariate Gaussian model and use the expectation maximization method to optimize the model parameters [18]. Considering multivariate Gaussian centered on the clusters, the likelihood of the data for the cluster $c_k$ is $p(x|c_k, \mu_k, \Sigma_k)$, where $x$, $\mu_k$ and $\Sigma_k$ are respectively the signals, mean and covariance matrix of cluster $c_k$, and $k$ is the number of clusters. Here, the number of clusters should also be predefined. An assumption for this clustering algorithm is that data are selected independently from the clusters. So, the marginal likelihood is not conditioned on the clusters and computes as

$$p(x|\theta_{1:K}) = \sum_k p(x|c_k, \theta_k)p(c_k);$$

$$\sum_k p(c_k) = 1, \theta_{1:K} = \{\mu_1, \Sigma_1, \dots, \mu_K, \Sigma_K\} \quad (5)$$

where $\theta_{1:K}$ defines the parameters for all the clusters and $p(c_k)$ is the probability of the cluster k and corresponds to the spikes firing frequencies [2]. Clustering performs by the calculation of probability of belonging data points to each of the clusters. This can be achieved by applying Bayes rule

$$p(c_k|x, \theta_{1:K}) = \frac{p(x|c_k, \theta_k)p(c_k)}{p(x|\theta_{1:K})} \quad (6)$$

It defines the Bayesian decision boundaries for the model. The class parameters optimize by maximizing the likelihood of the data as

$$\max p(X|\theta_{1:K}) = \max \prod_{n=1}^{N} p(x_n|c_k, \theta_{1:K}) \quad (7)$$

where $N$ is the number of vectors $X \equiv \{x_1, x_2, \ldots, x_N\}$.

Maximum likelihood techniques support models with more parameters, as the data can be better adjusted through extra parameters. However, the Bayesian model has a built-in trade-off between model complexity and the fit to the data [18].

*Fuzzy maximum likelihood estimation algorithm*

The fuzzy maximum likelihood estimation (FMLE) algorithm or Gath-Geva algorithm is proposed by Gath and Geva in 1989 [19]. FMLE is an extension of FCM algorithm and computes a fuzzy covariance matrix for each cluster. This makes the algorithm capable of recognizing elliptical clusters, as well as clusters with other shapes. The FMLE exponential distance measure-based algorithm is also able to detect clusters with variations in sizes, shapes and densities, even when there exist a large variety of them [28]. The FMLE clustering algorithm formulates according to [19], [29].

*F. Cluster Validity*

Since clustering algorithms recognize clusters which are not exactly predefined, the results of data clustering mostly require some sort of evaluation [30]. The validation procedure of clustering results is known as cluster validity methods.

Validity indices are extremely important tools to determine the optimal number of clusters. Since each cluster represents the firing activity of a single neuron, it is essential to accurately compute the optimal number of clusters based on each clustering method. Invalid clustering due to careless choice of cluster numbers result in inaccurate interpretation of neural activity of different brain areas. One serious consequence of this matter is ineffectiveness of neurological treatments due to misunderstanding of changes in neuronal activity caused by a certain disease. Also, it causes inaccuracies in simulations of the artificial neural networks based on the real neuronal activities.

In order to achieve the optimal number of clusters, first step is to estimate an upper bound and lower bound for the number of clusters and run the algorithm for each $c \in \{c_{min}, \ldots, c_{max}\}$ with $c$ be the number of clusters. Second step is to apply the validity function to obtain the optimal number of clusters based on the indirect analytical comparisons. There are different validity tools to evaluate each of algorithms. The techniques applied in this paper are discussed below.

*Xie-Beni index*

The Xie-Beni index [20], XB, calculates the overall average compactness against separation of the clusters. Compactness declares how close the cluster members are to each other. A common measure of compactness is the variance, which should be minimized. Separation shows how far the clusters themselves are from each other, that they should be widely scattered. The smaller values of XB represent compact and well-separated clusters. The minimized value of the index can be reached by the optimal number of clusters.

$$XB(c) = \frac{\sum_{k=1}^{n} \sum_{i=1}^{c} (\mu_{ik})^m \|x_k - v_i\|^2}{n \min_{i,k} \|x_k - v_i\|^2} \quad (8)$$

where $\mu_{ik}$ is the belonging probability of data point $k$ to cluster $i$ and $v_i$ is the center of cluster $i$.

*Calinski and Harabasz index*

Another index that we use for evaluation of the clustering algorithms is Calinski and Harabasz index [21], or CH index. This method determines the optimal number of clusters as

$$CH(c) = \frac{BCSS_c/(c-1)}{WCSS_c/(n-c)} \quad (9)$$

where BCSS is the between cluster sum of squares and WCSS is the within cluster sum of squares. Maximum value of this index represents the best splitting in dataset.

*Fuzzy hyper volume index*

The fuzzy hyper volume index, $V_{FH}$, which is introduced by Gath and Geva [19], is based on the concepts of hyper volume and density. $V_{FH}$ index considers the sum of all cluster sizes. A good partition should have a low fuzzy hyper volume, as small values indicate the existence of compact clusters. The fuzzy hyper volume $V_{FH}$ can be computed as $V_{FH} = \sum_{i=1}^{c} v_i$, where $v_i$ is the measure of cluster compactness [29]. Here, in addition to prior validity indices, the $V_{FH}$ validity index is used to find the optimal number of clusters in FMLE clustering algorithm.

Additionally, to evaluate the results of spike sorting via different algorithms, inter-spike interval (ISI) technique has been applied. This tool accumulates the time between spikes into inter-spike interval histograms.

## III. SIMULATION RESULTS

Considering the real basolateral amygdala data and using different spike sorting sub-techniques lead to several important results. Applying the three mentioned methods to set the threshold for spike detection, better results were achieved by performing the standard deviation-based technique. The value of $k$ was set to all three possible amounts for median and standard deviation-based methods and $k=4$ showed higher signal to noise ratio and used for further analysis. Applying manual threshold discrimination tool, Plexon offline sorter, we differentiated the amplitude distribution of spikes versus background activity. The performance of this method was also good. However, in order to achieve a completely automated analysis method, without involvement of the experimenter's bias, it is more preferable to apply fully automatic, mathematic-based techniques. The median-based method applied for spike detection was not capable of removing the proper amount of noise from the data. Applying this method, the detected spike waveforms were still contaminated with noise. So, the signals detected from threshold discrimination by the standard deviation-based method with $k=4$ was used for all clustering.

After detecting the spikes via threshold discrimination, we collected the spike waveforms for about 2 ms long, considering around 25 data points before the onset of detected spikes and 90 data points after that. These waveforms are the input for spike feature extraction and clustering algorithms.

Feature extraction was performed using both PCA and KPCA. Comparison of the clustering results based on PCA and KPCA indicated that KPCA is a much more reliable method for spike feature extraction than PCA, as it also considers the nonlinear features of the spikes. Applying Gaussian radial basis function Kernel PCA, better results were achieved by the optimal value $\beta = 10^{-3.85}$ for our dataset.

Clustering by most of clustering algorithms requires the number of clusters to be predefined. In the algorithms' implementation, the process starts by considering the value of minimum and maximum number of clusters. According to the results of performing the algorithm, the clustering index should be computed. The cluster number that makes the validity index to be optimal is the best. In this paper, the optimal number of clusters for all algorithms were obtained by applying XB and CH validity indices. The results of XB index for all algorithms and CH index for K-means and Bayesian algorithms are illustrated here. The lowest value of XB index denotes compactness and well-separation of the data, and consequently, the optimal cluster number. The optimal values of CH index is its maximum value which represents the best splitting in the dataset. $V_{FH}$ index was also applied to find the optimal number of clusters for FMLE clustering algorithm. Minimum value of $V_{FH}$ index indicates compact clusters and is the optimal one.

Fig. 1 and Fig. 2 represents the results of feature extraction conducted by PCA and KPCA for clustering via FCM algorithm, respectively. Comparing two plots, KPCA can segregate clusters from each other in a more reliable way.

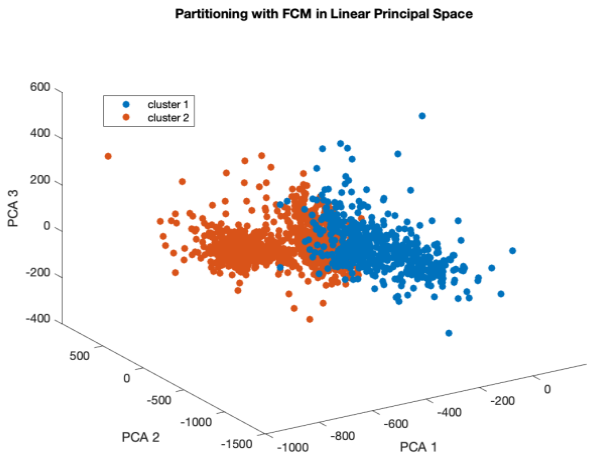

Fig. 1. Partitioning in data using PCA for FCM clustering algorithm.

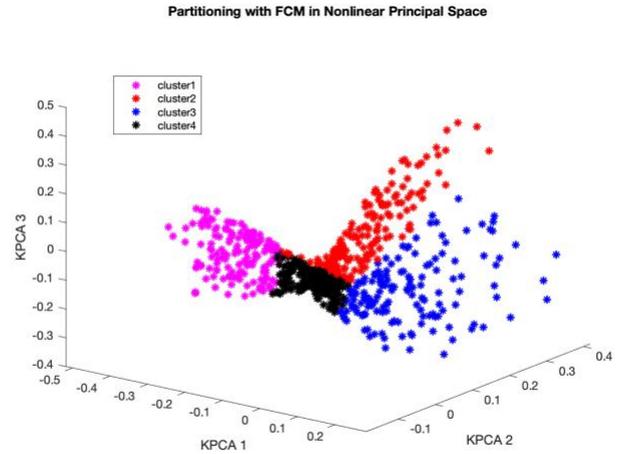

Fig. 2. Partitioning in data using KPCA for FCM clustering algorithm.

It is also noticeable that using KPCA technique for extracting features from a large dataset can result in elevating computational expenses as the kernel matrix is of the dimension $n \times n$ with $n$ being the number of data points. Still, for the dataset used in this study, consisting 10,000,000 data points, it works fast.

Fig. 3 and Fig. 4 represent the results of spike sorting according to a particular spike detection and clustering algorithm differentiating by the feature extraction tools, PCA and KPCA. As shown in Fig. 3, using PCA for extracting features of Fuzzy C-means algorithm results in two sorted clusters that are still contaminated by noise and integrated with activity of other neurons. Cluster 1 consists of spikes with different amplitudes that certainly should have not be fitted into one cluster. Second cluster is slightly better sorted than cluster 1. Still, it consists of different ranges of spike amplitudes. On the other hand, KPCA does a great job in extracting the spike features for FCM algorithm, shown in Fig. 4. According to plots, applying KPCA, four clusters consisting three well-sorted clusters are achieved. Cluster 4 shows some temporary delays which can emerge from the error of data acquisition system in capturing signals. Achieved results reveal the high capability of KPCA in extracting features and separating the signals.

The upper and lower bound of number of clusters for FCM algorithm were considered as $c_{min} = 2$ and $c_{max} = 12$ with $c$ being the number of clusters. Fig. 5 shows the results of applying XB validity index for FCM algorithm with KPCA as the feature extraction technique. As the smaller values of XB represent compact and well-separated clusters, applying this method, the best number of clusters is 4.

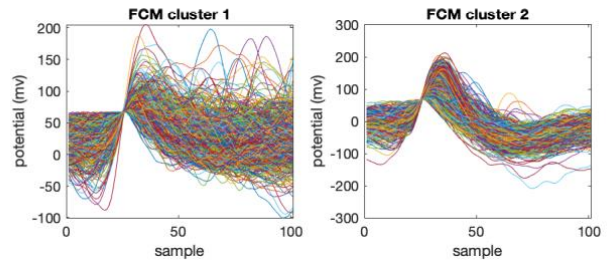

Fig. 3. Spike sorting via FCM algorithm clustering using PCA.

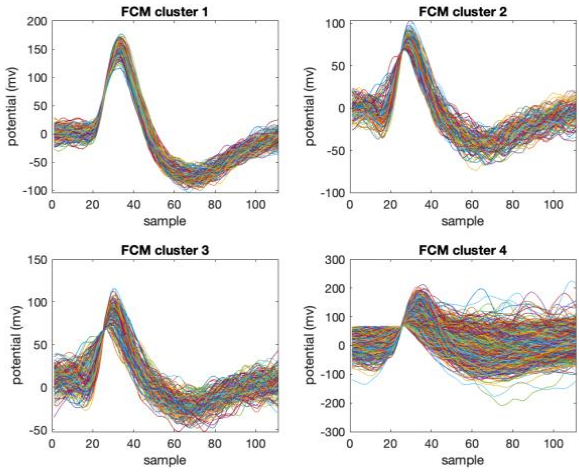

Fig. 4. Spike sorting via FCM algorithm clustering using KPCA.

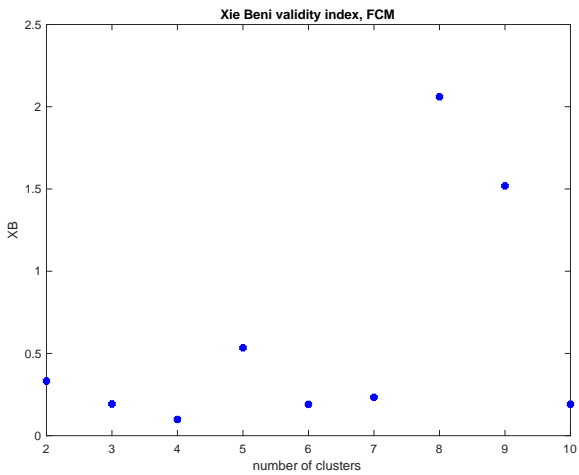

Fig. 5. XB index results for FCM clustering algorithm.

A measure for considering a cluster as a well separated one with good spikes is its spikes shape, repolarization, and depolarization, etc. Additionally, lower amounts of noise and artifacts, and similar spike timing result in a cleaner cluster. Also, a single neuron with good sorted spikes would have a great corresponded inter-spike interval histogram in terms of coefficient of variations and peaks.

As it is previously established, one technique to evaluate the quality of a sorted cluster is to plot the distribution or histogram of the inter-spike interval, between two consecutive spikes. The inter-spike interval histogram should be represented for all clusters to measure the regularity of spike timings within each of them. This histogram depicts the refractory period, i.e., a dearth of spikes that occur within milliseconds of each other. As we know, each cluster shows the activity of a single neuron and each neuron fires with a specific frequency. If spikes occur in regular intervals, the ISI histogram will have a sharp peak and coefficient of variations in ISI histogram will be small. Larger amplitude and existence of multiple peaks in ISI histogram represent the signals irregularity or artifact presence within that cluster. According to data using in this study, BLA neurons under normal condition, neurons fire with a regular rate. So, the inconsistency in ISI histogram is mostly a matter of sorting. Also, spikes preceding with very short intervals i.e., 3 ms convey information most efficiently. Number of bins for ISI plots of all applied algorithms is set to 200.

Fig. 6 and Fig. 7 show the ISI histogram for the cluster of spikes sorted via FCM algorithm represented in Fig. 3 and Fig 4. Fig. 6 shows several peaks with high amplitudes. It can be inferred that the neural activity of each of the two clusters is related to firing of more than one neuron which is somewhat expected from the analysis of Fig. 3. However, using KPCA as the tool for feature extraction, ISI histograms represent better sorted clusters. Comparing the ISI plots in Fig. 6 and Fig. 7 further illustrates the limitations of PCA for extraction of spike features and capability of KPCA in signals segregation.

In the following, we discuss the results of spike sorting applying the rest of algorithms, i.e. K-means algorithm, Bayesian clustering algorithm, and FMLE algorithm, considering KPCA as the feature extraction tool. Although the ISI histogram in also plotted to evaluate the clusters of these three algorithms, they are not illustrated in the paper.

K-means clustering starts by initializing a set of $k$ cluster centers. Afterwards, it assigns each datum to the cluster whose center is the nearest, and calculates the centers again. This process continues until a halt in changes of the cluster centers happens. The upper bound and lower bound for number of clusters is set as $c_{min} = 2$ and $c_{max} = 10$. The results of applying K-means algorithm and the validation outcomes are shown in Fig. 8 and Fig. 9. According to both validity indices, the optimal number of clusters for this algorithm is 4.

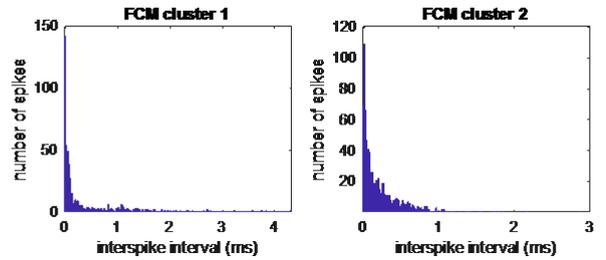

Fig. 6. Inter-spike interval histogram for FCM algorithm using PCA.

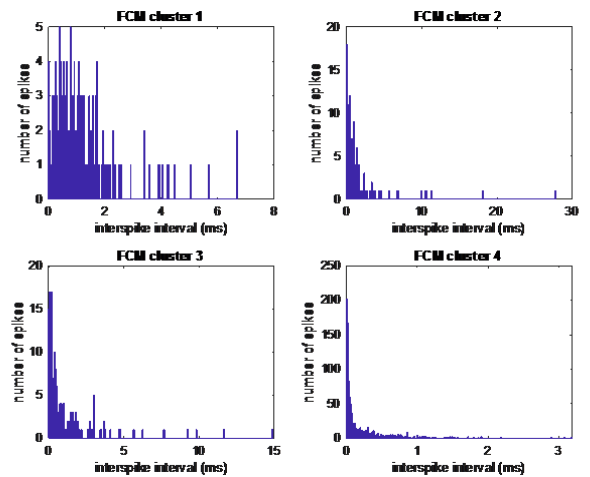

Fig. 7. Inter-spike interval histogram for FCM algorithm using KPCA.

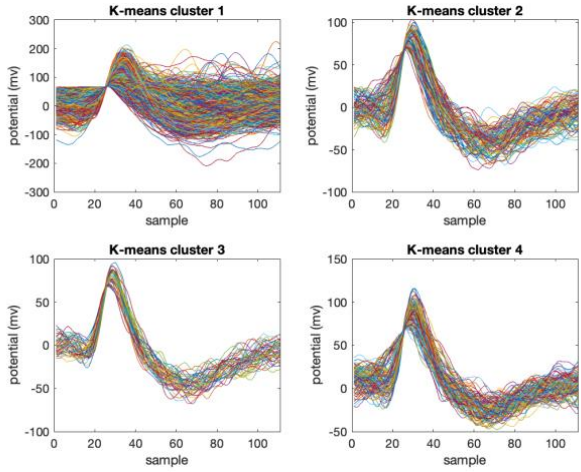

Fig. 8. Spike sorting via K-means algorithm using KPCA.

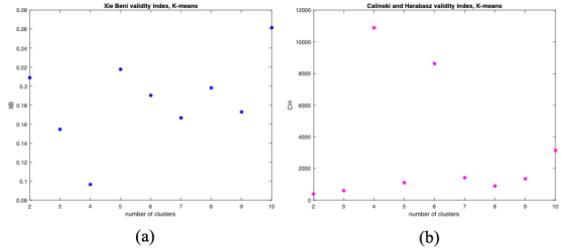

Fig. 9. CH and XB indices results for K-means algorithm.

It should be in mind that although K-means algorithm is a simple and popular clustering method among neuroscientists, it has some drawbacks. The major problem of k-means clustering is that it works well in capturing structure of the data only if clusters have a spherical shape. In other words, K-means clustering algorithms is unable to capture clusters structure if they have a complex geometric shape. Additionally, this algorithm prevents data points which are far from each other to be in one cluster, even if they indeed belong to one cluster.

Fig. 10 and Fig. 11(a), (b) represent the results of spike sorting via Bayesian clustering algorithm. The upper bound and lower bound for number of clusters are considered as $c_{min} = 2$ and $c_{max} = 16$. To determine the optimal number of clusters based on this algorithm, validity indices consisting XB and CH indices are applied. The results of validations can be seen in Fig. 11 (a), (b). As it is shown, based on XB index, 2 to 5 clusters are fine and based on CH index 5 is the optimal number of clusters. So, we consider $c = 5$ for this clustering algorithm. One advantage of Bayesian clustering is the possibility of quantifying the certainty of clustering. It is helpful in deciding about the spikes' isolations in different clusters. Exploring the probability distribution of clusters achieved by (6) gives a quality measure of each cluster' separation.

Fig. 11(c), (d) and Fig. 12 illustrate the results of spike sorting with FMLE algorithm. To achieve a better result with FMLE algorithm, we begin the process by applying a K-means algorithm. The results are employed as the initial cluster centers for FMLE algorithm. The weighting term $m$ which expresses the fuzziness of each cluster in FMLE clustering algorithm is set to $m = 3$ and the number of iterations considered as 70. The upper and lower bound of cluster numbers are set to $c_{min} = 2$ and $c_{max} = 6$ and the optimal number of clusters are $c = 4$, applying validity indices shown in Fig. 11 (c), (d).

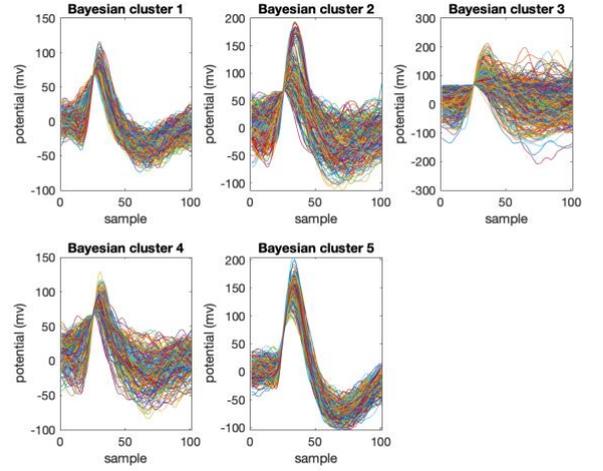

Fig. 10. Spike sorting via Bayesian clustering algorithm using KPCA.

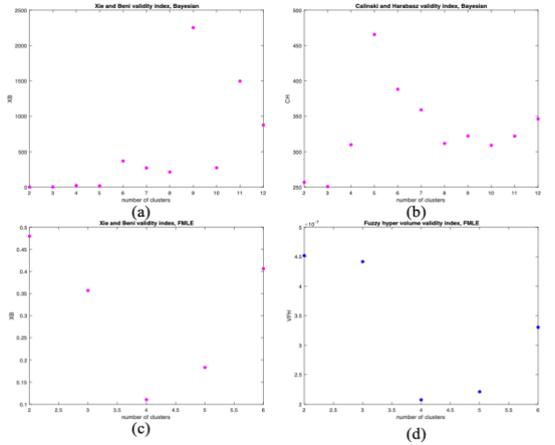

Fig. 11. Validity results for Bayesian clustering and FMLE algorithm.

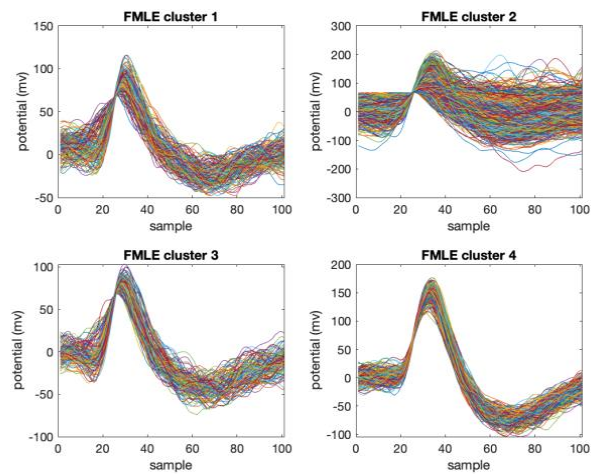

Fig. 12. Spike sorting via FMLE clustering algorithm using KPCA.

## IV. CONCLUSION

This paper is a systematic experimental comparison of different spike sorting methods employing data from the rat brain. To represent the differences in results associated with applying different spike sorting sub-techniques based on the experimental data, we have applied various methods. For threshold discrimination, we have performed three methods consisting traditional practical method, median-based method, and standard deviation-based technique. Among these methods, the standard deviation-based approach outperformed. PCA and KPCA have been adopted to extract features from spike waveforms. Comparison of data partitioning based on PCA and KPCA have indicated that KPCA is a robust and more reliable method for spike feature extraction than PCA. Spike sorting have been completed employing different clustering algorithms including K-means, Fuzzy C-means, Bayesian and Fuzzy maximum likelihood estimation. Although K-means algorithm is a fairly simple and popular clustering algorithm among neuroscientist, its performance was lower than the rest. Among applied clustering tools, fuzzy methods i.e. FCM and FMLE performed better and represented higher quality clusters. Evaluation have been done applying three methods consisting XB, CH, and $V_{FH}$ to achieve the optimal number of clusters. At the end, the quality of each cluster is evaluated via the inter-spike interval histograms. Applying mentioned methods, we have achieved the segregated activity of single BLA neurons. The results provide a basis for neuroscientists to choose effective spike sorting sub-methods in future real data analysis. This will provide a basis for a better diagnosis and treatment of neurological disorders. Achieved results strongly suggest to employ the standard deviation-based method for threshold discrimination, apply KPCA for feature extraction, and perform FCM or FMLE as the clustering algorithm for spike sorting.